\documentclass[pre,twocolumn,showpacs
]{revtex4}
\newcommand{\be}{\begin{equation}}
\newcommand{\ee}{\end{equation}} 
\newcommand{\bea}{\begin{eqnarray}} 
\usepackage{graphicx}
\newcommand{\eea}{\end{eqnarray}}
\usepackage{epsfig}
\usepackage{bm}

\begin{document}
\title{A Minimalist Turbulent Boundary Layer Model}
\author{L. Moriconi}
\affiliation{Instituto de F\'\i sica, Universidade Federal do Rio de Janeiro, \\
C.P. 68528, 21945-970, Rio de Janeiro, RJ, Brazil}
\begin{abstract}

We introduce an elementary model of a turbulent boundary layer over a flat surface, given as a vertical random distribution of spanwise Lamb-Oseen vortex configurations placed over a non-slip boundary condition line. We are able to reproduce several important features of realistic flows, such as the viscous and logarithmic boundary sublayers, and the general behavior of the first statistical moments (turbulent intensity, skewness and flatness) of the streamwise velocity fluctuations. As an application, we advance some heuristic considerations on the boundary layer underlying kinematics that could be associated with the phenomenon of drag reduction by polymers, finding a suggestive support from its experimental signatures.

\end{abstract}
\maketitle

Turbulent boundary layers have been a central topic of interest in fluid dynamic research for long years 
\cite{bl100,schli}. Nevertheless their obvious technological importance, a satisfactory description of the physical mechanisms which underlie the boundary velocity fluctuations remains elusive to date. As the result of intensive computational and experimental efforts carried out mainly along the last two decades, it is by now clear that the turbulent boundary layer is the stage for the production and the complex interaction of coherent structures \cite{rob}, a fact that was formerly antecipated by Theodorsen \cite{theo} and Townsend \cite{town}.

Standard phenomenological formulations of the turbulent boundary layer problem aim at solving self-consistent equations for the expectation values of velocity and the Reynolds stress tensor components, relevant quantities in engineering applications \cite{schli,pope}. At the very conception of these models no fundamental role is given to the whole boundary layer zoo of coherent structures, like streamwise and hairpin vortices, low speed streaks, etc. It is an open question, for instance, if a structural derivation of the Prandtl von-Karman logarithmic law of the wall is viable. In this sense, turbulent boundary layer modelling is a difficult problem of statistical physics, analogous to the derivation of thermodynamic equations of state from molecular kinematics/dynamics. The literature on the subject is still relatively small, although an initial discussion may be traced back to 1982 with Perry and Chong \cite{perry-chong}.

This work differs from previous attempts \cite{perry-chong, perry2, sree, perry3, marusic1,marusic2} essentially in its stronger simplifying assumptions, specific coherent structure modelling (the vortex-dipole model to be introduced below), and the analysis of higher order statistics for the streamwise velocity fluctuations. We do not seek, at the present level of mathematical treatment, quantitative agreement with experiments; instead, we compute a set of general profiles, which turn out to be clearly supported by observations.

Our focus is on the streamwise fluctuations of the velocity field. Let us assume that close enough to the wall these fluctuations are mostly due to the flow generated by hairpin vortices \cite{head,wu,adrian}, like the one depicted in Fig.1, momentarily located in the surroundings of the measurement position. The main contribution to streamwise fluctuations would come from the spanwise sector of hairpin vortices (also called ``hairpin's head" ), while subdominant contributions would be related to their necks and legs. 
%\vspace{0.1cm}

%%
\begin{center}
\begin{figure}[tbph]
\includegraphics[width=6.0cm, height=6cm]{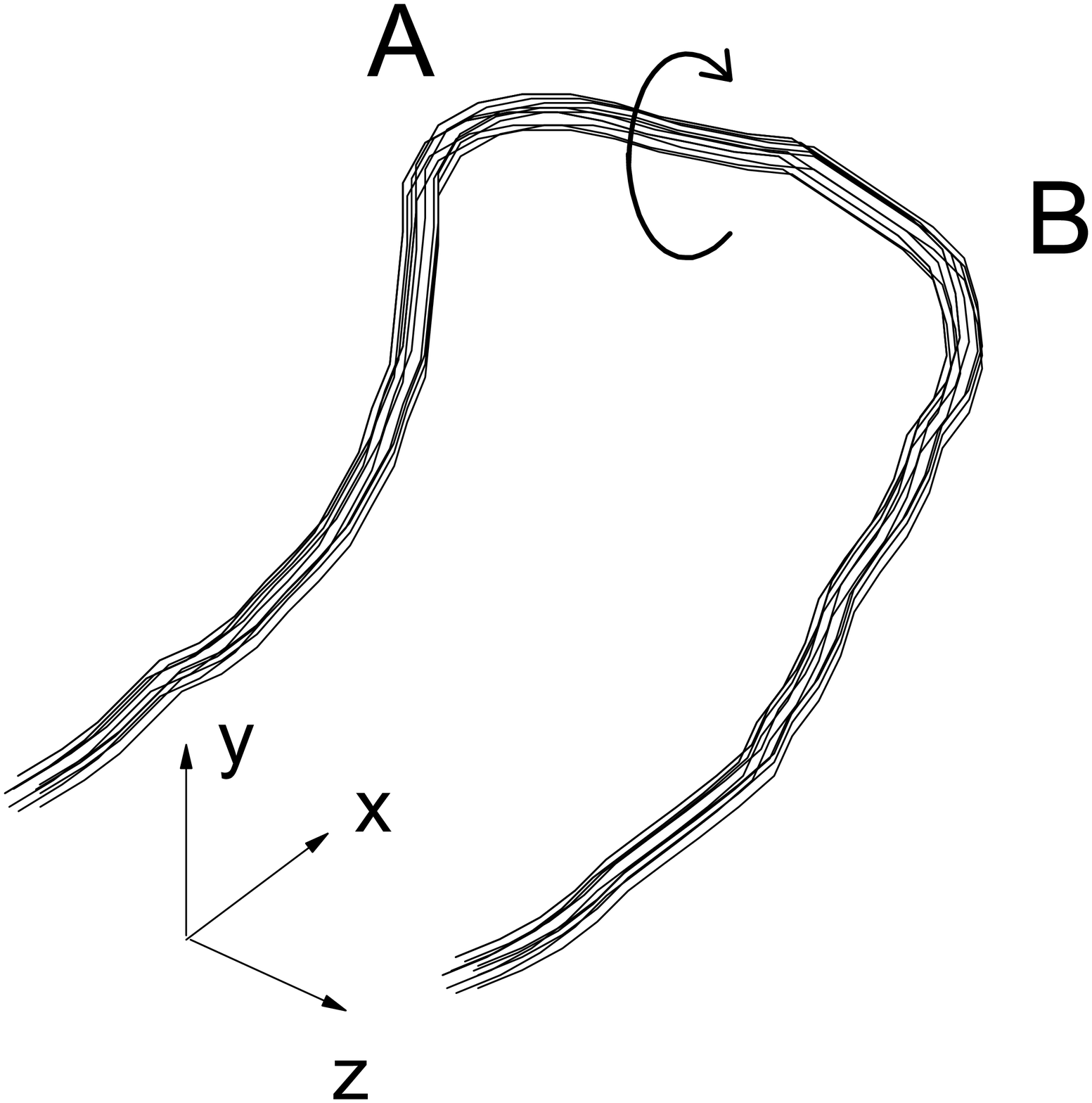}
\caption{A hairpin vortex which propagates along the positive $x$ direction. Streamwise fluctuations of the velocity field, associated to the passage of the hairpin vortex, are due essentially to the flow generated by its spanwise sector, which lies between points A and B. The curved arrow indicates the vorticity orientation.}
\label{fig1}
\end{figure}
\end{center}
\vspace{-0.5cm}

Since streamwise fluctuations of velocity are the only ones we have in mind, it is natural to replace the hairpin vortex by a simpler and more mathematically tractable structure, which we take to be a spanwise Lamb-Oseen vortex, an exact and non-stationary solution of the Navier-Stokes equations, described in cilindrical coordinates as
\be
u_\theta (r) = \frac{\Gamma}{2 \pi r} [1 - \exp(-r^2/2r_c^2)]  \label{lamb-oseen} \ , \
\ee
where $\Gamma$ is the total circulation around the vortex and $r_c^2=2 \nu t$ is the squared vortex core radius at time $t$, defined in terms of the kinematical viscosity $\nu$.

Of course, (\ref{lamb-oseen}) solves the fluid equations of motion in the absence of boundaries, so (\ref{lamb-oseen}) is just a rough approximation to a real vortex parallel to the wall. In our modelling definitions, we postulate that the symmetry axis of the vortex lies in the plane (henceforth designed the ``measurement plane") that contains the measurement point and is normal to the wall. We assume, then, that at equally spaced time intervals, a given vortex is replaced by another one, at a random distance $y$ from the wall, with some prescribed probability distribution function (pdf) $\rho(y)$. 
\vspace{0.1cm}

\begin{center}
\begin{figure}[tbph]
\includegraphics[width=6.4cm, height=6cm]{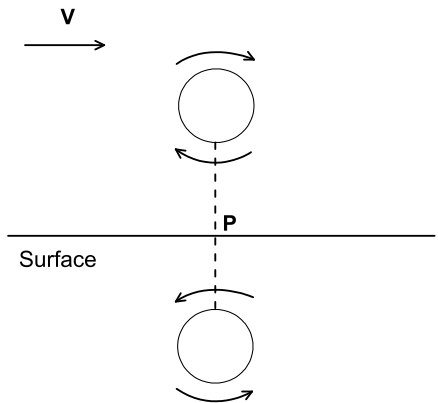}
\caption{The vortex dipole construction. The dashed line is contained in the measurement plane. The upper plane vortex is the ``real" one, while the other is its mirror image. A uniform backgroung flow with velocity $V$ is superimposed to the velocity field produced by the vortices, so that $P$ be
a stagnant point.}
\label{fig1}
\end{figure}
\end{center}
\vspace{-0.5cm}

As the model under consideration is effectively two-dimensional, a mirror vortex is introduced ``below the wall", so that streamlines do not cross the material surface. Furthermore, in order not to completely neglect the non-slip boundary condition, some improvement is attained if we make the velocity field to vanish at the intersection of the measurement plane with the wall. For this purpose, an external homogeneous velocity field is superimposed to the field generated by the vortex dipole. These definitions are shown in Fig. 2.

An interpretation of the time dependence in Eq. (\ref{lamb-oseen}) is in order. The time variable $t$ is taken to be the total time elapsed since the hairpin vortex was created at the wall. This assumption, however, is not of great help, if there is no way to relate the vortex vertical position $y$ to the time $t$. To solve this problem, at least in a phenomenological fashion, we find inspiration in the scaling structure of the {\it{laminar}} boundary layer over a flat surface. It is well-established, and analytically predicted by the Blasius solution, that the laminar boundary layer thickness grows with the distance from the leading edge as $\delta \sim \sqrt{x}$. This result can be understood in elementary terms as the fact that any small perturbation which is transported along the main direction of the flow (say, the horizontal one) follows a diffusive drift along the vertical direction. An analogy to the context of turbulent boundary layers can be drawn, replacing words like ``perturbations" by ``coherent structures" and ``molecular viscosity" by ``eddy viscosity". Actually, hairpin vortices are observed to grow in size as they get farther from the surface 
\cite{adrian}. Therefore, we suppose that at time $t$, a diffusion-like relation $t \sim y^2$ holds, and the vortex core radius can be written as $r_c = a y$ in (\ref{lamb-oseen}), where $a$ plays the role of a phenomenological parameter in the turbulent boundary layer modelling \cite{comment}.

The flow depicted in Fig. 2 is assumed to describe a boundary layer with no pressure gradient. This is so because the velocity field is symmetric under reflections on the measurement plane. Pressure gradients are probably related to the flow induced by the whole gas of hairpin's vortices, appearing, therefore, as a ``many-body" effect.

We are now ready to work out a few relevant equations. Suppose that the Lamb-Oseen vortices are centered at $y$ and $-y$. The streamwise velocity field vanishes at point $P$ in Fig. 2. Once the resulting streamwise velocity is given as the sum of three contributions (the external, the real and the mirror vortex velocity fields), we get, using Eq. (\ref{lamb-oseen}),
\be
0 = V - \frac{\Gamma}{\pi y} [1 - \exp(-1/2a^2)] \ . \ \label{vel-p}
\ee
The above equation holds for any $y$ only if the total circulation $\Gamma$ is a $y-$dependent quantity. We find
\be
\Gamma(y) = \frac{\pi V y}{1 - \exp(-1/2a^2)} \ . \
\ee
It is a straightforward task to evaluate expectation values of general $y-$dependent functionals $F=F[u(y)]$ of the streamwise velocity field. It follows that
\be
\langle F[u(y)] \rangle = \int_0^\infty dy' \rho(y') F[u(y,y')] \ , \ \label{av}
\ee
where the ``two-point velocity",
\bea
u(y,y') &=& V + \frac{\Gamma(y') }{2 \pi (y-y')} [1 - \exp(-(y-y')^2/2a^2 y'^2)] \nonumber \\
&-& \frac{\Gamma(y') }{2 \pi (y+y')} [1 - \exp(-(y+y')^2/2a^2 y'^2)] \ , \
\eea
is nothing but the velocity field at $y$ when the Lamb-Oseen vortices are placed at $y'$ and $-y'$.

We have applied (\ref{av}) for a set of velocity functionals, comparing the $y-$dependent profiles so obtained with experimental results, as discussed below. As input parameters, we take $a=1.0$, $V=1.0$ (i.e., the streamwise velocity is computed in units of the external velocity $V$). We use the pdf 
\be
\rho (y) = \frac{2b}{\pi (y^2+b^2)} \label{lor} \ , \
\ee
with $b=1.0$, to model fluctuations of the vortex's height above the surface. It is important to note that the choice of the lorentzian distribution (\ref{lor}), although arbitrary, is by no means restrictive. We have checked out that smoothly decaying distributions lead to similar conclusions, if one is indeed interested in a {\it{qualitative}} understanding of turbulent boundary layer fluctuations.
\vspace{0.2cm}

{\leftline{$\bullet$ {\it{The viscous and logarithmic layers}}}}
\vspace{0.2cm}

The mean streamwise velocity is obtained from the expectation value of $F[u(y)]=u(y)$. The overall profile is shown in Fig. 3, which interpolates between zero velocity at the wall and unit velocity at infinity. Even though $u(y)$ seems to give a reasonable profile for the interval $0 \leq y < \infty$, the model does not apply to the outer layer, because of the stronger interactions between coherent structures and also for the high intermittency produced by the random entrainment of external laminar flow that take place in that region.

\begin{center}
\begin{figure}[tbph]
\includegraphics[width=9.2cm, height=7.0cm]{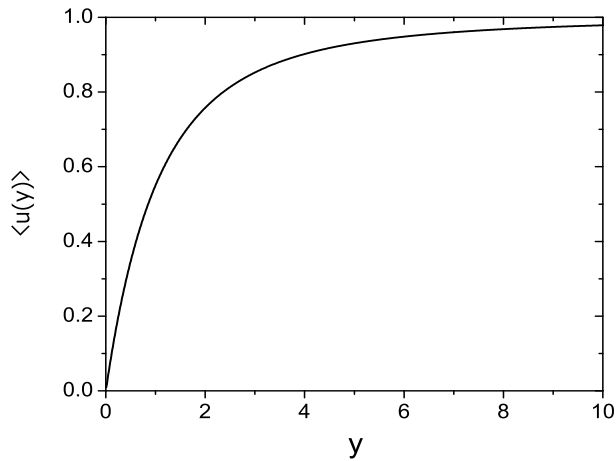}
\caption{The mean streamwise velocity.}
\label{fig3}
\end{figure}
\end{center}

\begin{center}
\begin{figure}[tbph]
\includegraphics[width=9.2cm, height=7.0cm]{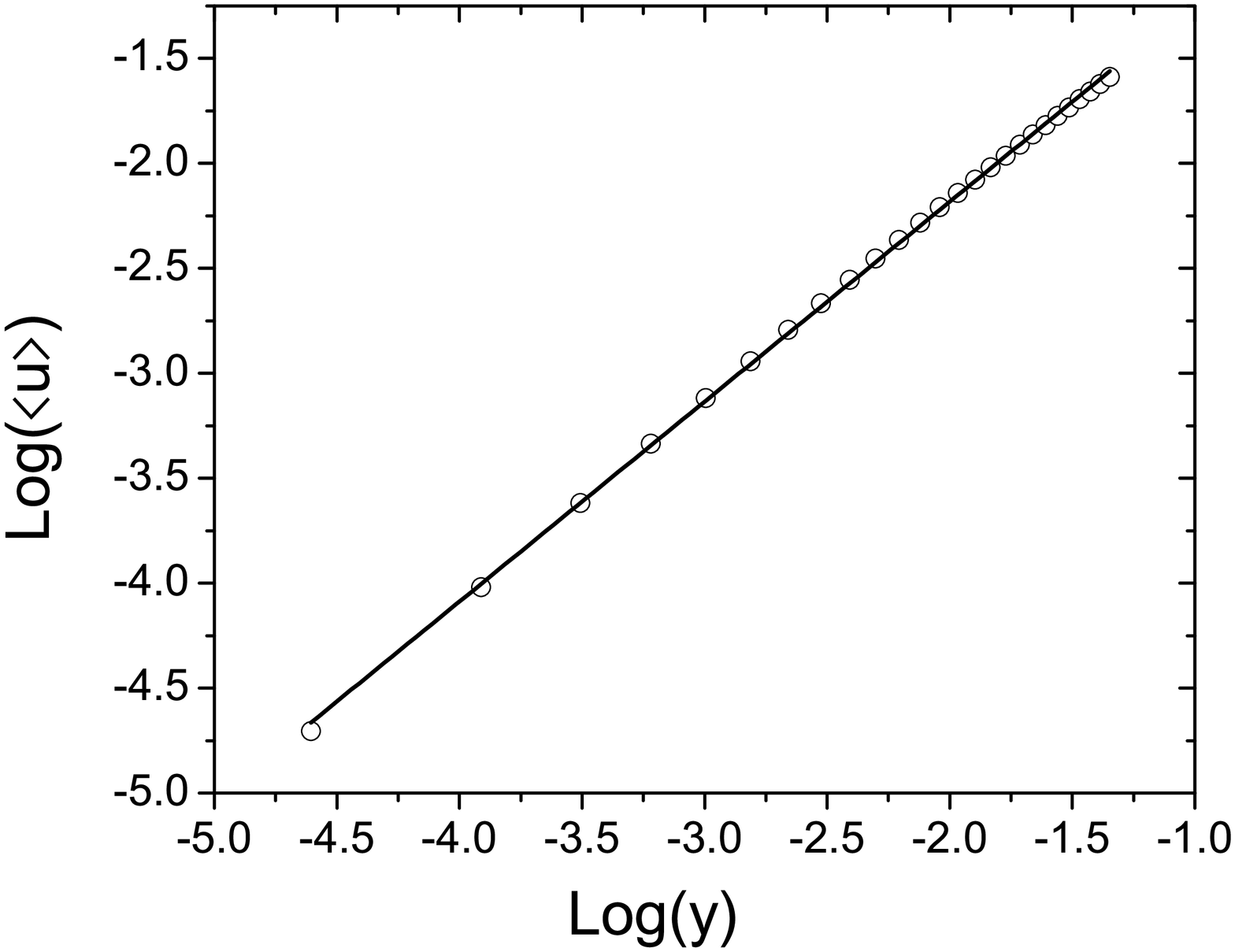}
\caption{The viscous layer, verified in the range 
$0.01 \leq y \leq 0.25$. The straight line has slope 0.95.}
\label{fig4}
\end{figure}
\end{center}
\vspace{-1.3cm}

In Figs. 4 and 5, viscous and log-layers are clearly noticed for certain ranges of vertical distances. The excellent fit of the data to the straight line in Fig. 5 is given by $\langle u(y) \rangle = 0.31 \ln(y) + 0.55$. The numerical coefficients have precision of 0.1\%. Observe that a purely dimensional argument yields $u(y) = V f(y/b)$ in our particular model. Therefore, the numerical verification of a log-layer forces us to identify the effective external velocity $V$ to the friction velocity, up to numerical factors. We may conjecture, thus, that the friction velocity is ``what is left" when a few dominant vortical structures near the wall are removed. In other words, the friction velocity can be interpreted here as a ``mean field", while fluctuations are modeled by isolated vortex dipoles.

\begin{center}
\begin{figure}[tbph]
\includegraphics[width=9.2cm, height=7.0cm]{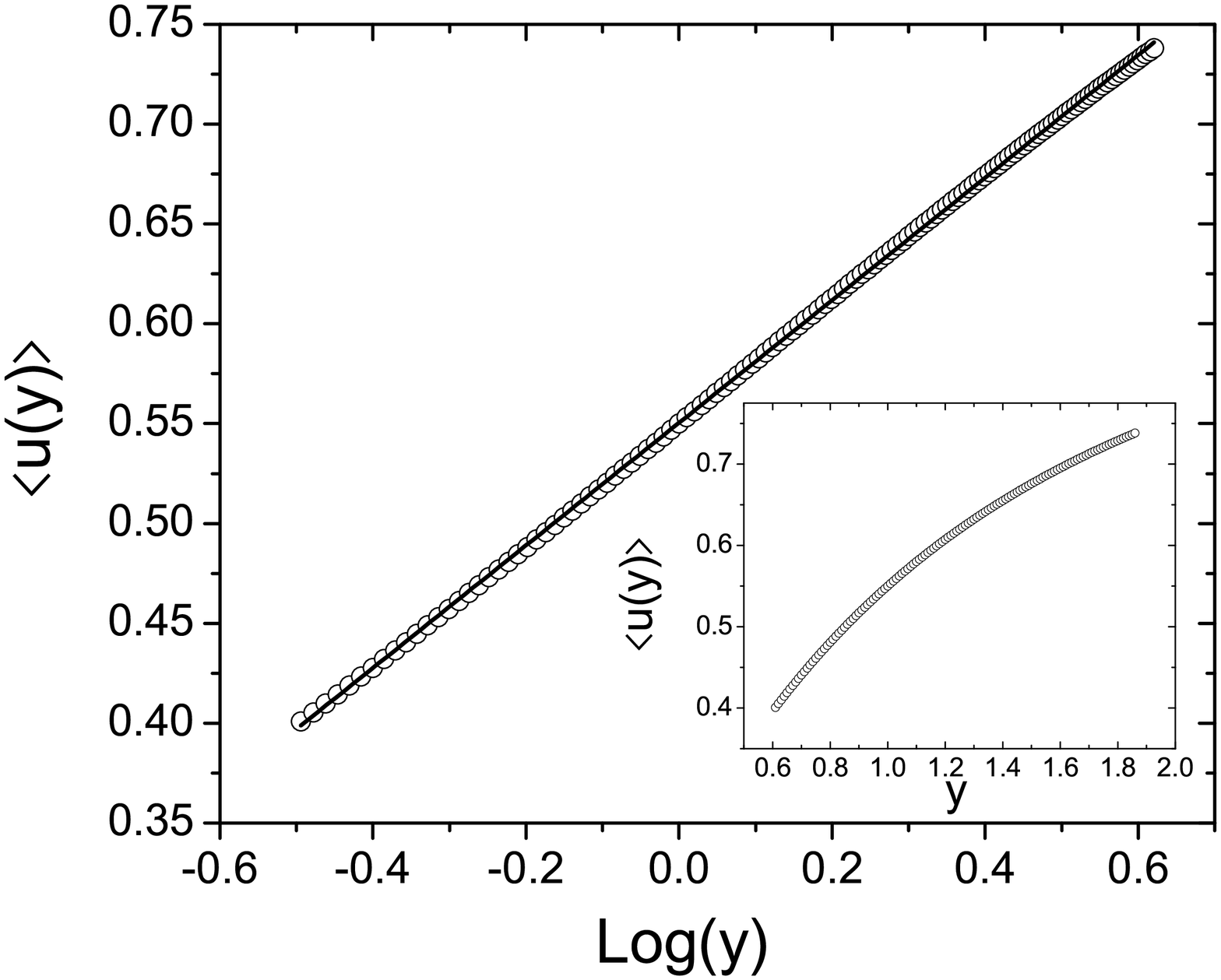}
\caption{The logarithmic layer, verified for $0.6 \leq y \leq 1.85$. The inset shows the same data plotted 
in linear scales.}
\label{fig5}
\end{figure}
\end{center}
\vspace{-0.5cm}

{\leftline{$\bullet$ {\it{Higher order statistics}}}}
\vspace{0.2cm}

Let $F_n[u(y)] \equiv [u(y)- \langle u(y) \rangle]^n$. We introduce
\be
u_{rms} = \sqrt{ \langle F_2[u(y)] \rangle }
\ee
and the hyperflatness functions $S_n(y)$, given by
\be
S_n(y) = \frac{\langle F_n[u(y)] \rangle}{\sqrt{\langle F_2[u(y)] \rangle^n}} \ . \
\ee
Fine measurements of turbulent boundary layer fluctuations for $S_3$ (skewness) and $S_4$ (flatness or kurtosis), which can resolve the region very close to the surface, are reported in Ref. \cite{lork}. As we can see from Figs. 6, 7 and 8, there is a clear qualitative agreement with observations. In particular, the abrupt sign-changing transition of skewness is remarkably reproduced by the vortex dipole model.

\begin{center}
\begin{figure}[tbph]
\includegraphics[width=9.2cm, height=7.0cm]{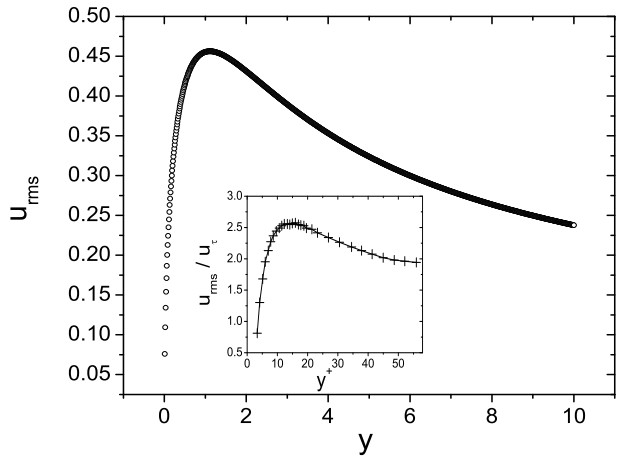}
\caption{The $u_{rms}$ velocity. The inset shows the experimental measurements of $u_{rms}$ by Lorkowski \cite{lork}.}
\label{fig6}
\end{figure}
\end{center}
\vspace{-1.2cm}

\begin{center}
\begin{figure}[tbph]
\includegraphics[width=9.2cm, height=7.0cm]{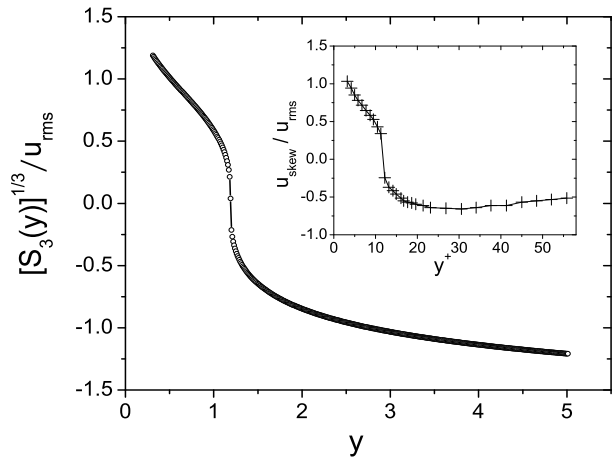}
\caption{The skewness for $0.3 \leq y \leq 5.0$. The inset shows the experimental measurements of $S_3(y)^{1/3}/u_{rms}$ by Lorkowski \cite{lork}.}
\label{fig7}
\end{figure}
\end{center}

\begin{center}
\begin{figure}[tbph]
\includegraphics[width=9.2cm, height=7.0cm]{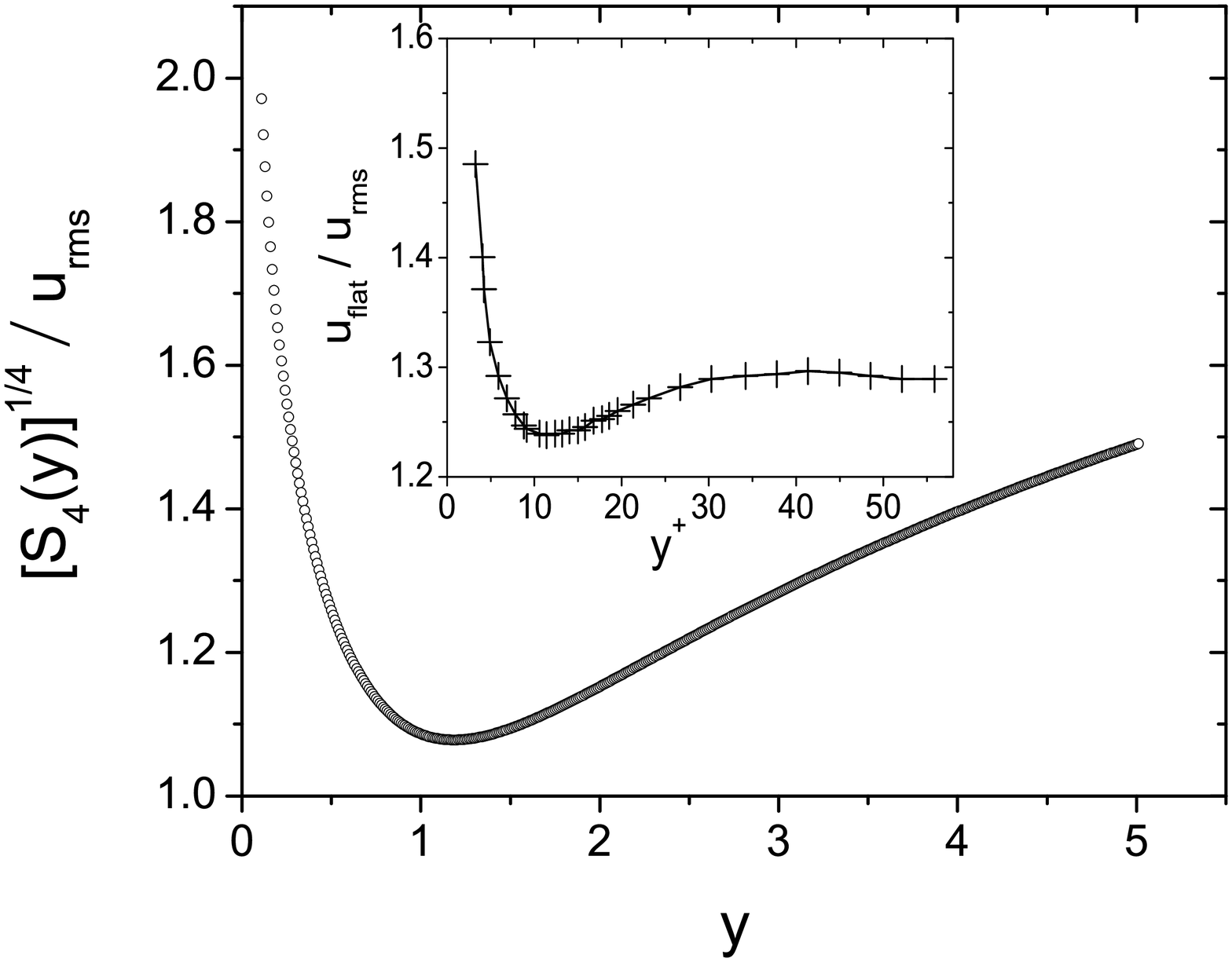}
\caption{The flatness for $0.1 \leq y \leq 5.0$. The inset shows the experimental measurements of $S_4(y)^{1/4}/u_{rms}$ by Lorkowski \cite{lork}.}
\label{fig8}
\end{figure}
\end{center}
\vspace{-1.5cm}

{\leftline{$\bullet$ {\it{Drag reduction by polymers}}}}
\vspace{0.2cm}

The phenomenon of drag reduction by polymers \cite{toms,proc} is a long-standing puzzle of non-newtonian fluid mechanics. The broad picture is that dissipation is attenuated due to the interaction of polymers with the coherent structures created near walls. 

Recent PIV (particle image velocimetry) experiments in flows with dilute polymers have shown that vorticity fluctuations - and probably coherent structures - are supressed at some point above the surface in turbulent boundary layers \cite{white}. Also, by about the same time, interesting signatures of drag reduction have been found in the profiles of $u_{rms}(y)$ and $S_3(y)$ in connection with polymer dilution \cite{itoh}. An additional peak is observed for $u_{rms}(y)$, while the skewness $S_3(y)$ shows further sign-changing transitions.

\begin{center}
\begin{figure}[tbph]
\includegraphics[width=9.2cm, height=7.0cm]{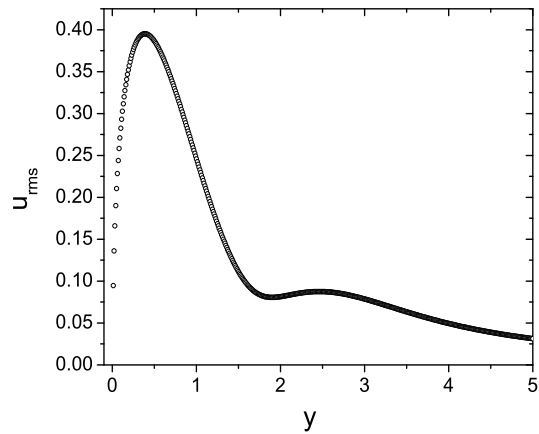}
\caption{The $u_{rms}$ velocity, as affected by coherent structure suppression, for $0.01 \leq y \leq 5.0$.
An additional peak is observed at $y \simeq 2.5$.}
\label{fig9}
\end{figure}
\end{center}

\begin{center}
\begin{figure}[tbph]
\includegraphics[width=9.2cm, height=7.0cm]{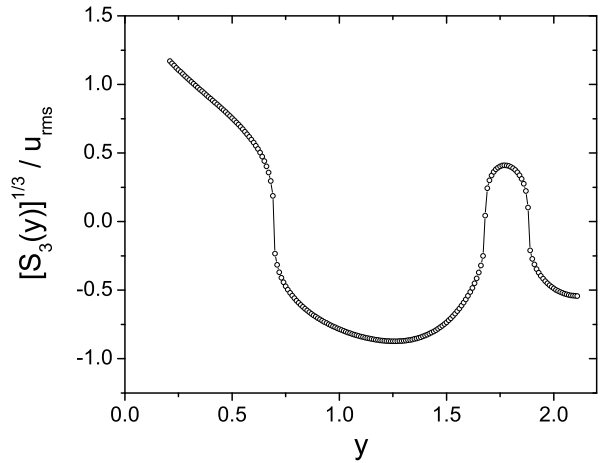}
\caption{The skewness, as affected by coherent structure suppression, for $0.2 \leq y \leq 2.1$. Note the additional sign-changing
transitions that take place within $1.5 < y < 2.0$.}
\label{fig10}
\end{figure}
\end{center}
\vspace{-1.0cm}

The vortex dipole model allows us to relate these apparently distinct features of drag reduction by polymers. Coherent structure supression can be naturally accounted for by changing the form of the vortex pdf $\rho(y)$. We take, for instance, a distribution which is uniform for $0 \leq y \leq c$, but vanishes for $y>c$, that is,
\be
\rho(y) = c^{-1}[\theta(y)-\theta(y-c)] \ , \
\ee
where $\theta(y)$ is the Heaviside function. Therefore, we suppose no vortex is found for $y > c$, as the result of polymer interactions. Setting $c=1.0$ and keeping the previous definitions of $a$, $b$ and $V$, we get the results shown in Figs. 9 and 10, which are in striking correspondence with real profiles \cite{itoh}.

We conclude with some general remarks. The present model does not take into account further aspects of the turbulent boundary layer phenomenology, which become important if the interest is shifted toward quantitative comparisons with experiments. An essential improvement, along the above lines, would be to introduce a pair of streamwise vortex configurations with opposite vorticity as a way to mimick hairpin legs. Only in this way it would be possible to compute the shear stress and, thus, define the physical scales of length and velocity which are necessary for the description of the inner boundary layer. 

%Alternatively, we may write down two-dimensional evolution equations for the vortex pdf (now taken as a function of $x$, %$y$ and $t$), assuming that the vortices are passively advected by the self-induced velocity field. We have, neglecting %streamwise velocity corrections,
%\be
%\frac{\partial \rho}{\partial t} + V \frac{\partial \rho}{\partial x}
%+  \frac{\partial (v \rho)}{\partial y} = 0 \ , \ \label{rho-eq} 
%\ee
%Where $v(y)$ is the $y-$component of velocity at the head of the hairpin's vortex. The asymptotically stationary %solutions of (\ref{rho-eq}) would give, in principle, important information about the boundary layer morphology. As %discussed above, in a simple extension of the vortex dipole model, hairpin's legs can be modeled by anti-parallel %Lamb-Oseen vortices, leading to $v(y) \sim V$ for $y$ large enough. The stationary solution of (\ref{rho-eq}) is, then, %of the form $\rho(x,y) = f(x-y)$, which indicates that the boundary layer thickness growth is given by $\delta \sim x$. %That is not a too bad result, if compared to the empirical law $\delta \sim x^{4/5}$ \cite{pope}.

To summarize, we have introduced an elementary model of a turbulent boundary layer, focusing our attention on the streamwise fluctuations of velocity induced by hairpin vortices. The model's main scope is to provide qualitative insights on the velocity and hyperflatness profiles. Up to the knowledge of the author, this is the first time the profiles of skewness and flatness of usual turbulent boundary layers, as well as certain statistical signatures of the phenomenon of drag reduction by polymers have been theoretically reproduced by means of vortex methods.

\acknowledgements
This work has been partially supported by CNPq and FAPERJ. I thank Atila Freire for several interesting discussions and for calling my attention to Refs. \cite{perry-chong,perry2,sree,perry3,marusic1}. I also thank Katepalli Sreenivasan for kindly providing me a copy of Ref. \cite{sree}.
%\end{acknowledgements}


\begin{references}
\bibitem{comment} There is no strong argument to rule out a general anomalous diffusive ansatz, $r_c = a y^\alpha$. We would not find, however, important qualitative departures for the results derived here.
\end{references}
\end{document}